\documentclass[review]{elsarticle}
\usepackage{amsmath,amssymb}
\newcommand{\be}{\begin{equation}}
\newcommand{\ee}{\end{equation}}
\newcommand{\bea}{\begin{array}}
\newcommand{\ea}{\end{array}}
\newcommand{\beqa}{\begin{eqnarray}}
\newcommand{\eeqa}{\end{eqnarray}}
\newcommand{\bean}{\begin{eqnarray*}}
\newcommand{\eean}{\end{eqnarray*}}
\begin{document}

\begin{frontmatter}



\title{Lattice vortices induced by noncommutativity}


\author[1]{A.A. Minzoni}
\ead{tim@mym.iimas.unam.mx}
\author[2]{L.R. Ju\'arez}
\ead{roman.juarez@nucleares.unam.mx}
\author[2]{M. Rosenbaum\corref{cor1}}
\ead{mrosen@nucleares.unam.mx}
\address[1]{FENOMEC-IIMAS, Universidad Nacional Aut\'onoma de M\'exico}
\address[2]{Instituto de Ciencias Nucleares-FENOMEC, Universidad Nacional Aut\'onoma de M\'exico}
\cortext[cor1]{Corresponding author}
\date{\today}

\begin{abstract}
We show that the Moyal $\star$-product on the algebra of fields induces an effective
lattice structure on vortex dynamics which can be explicitly constructed using recent
asymptotic results.
\end{abstract}


\end{frontmatter}




\section{Introduction}

The study of the behavior of classical fields defined as functions of noncommutative spatial variables has received a great deal of attention in the last few years (see {\it e.g.} \cite{gopa1},\cite{jack},\cite{bak},\cite{gopa2},\cite{tong} and \cite{olaf} for a review). In particular the study of coherent structures in the form of noncommutative solitons or noncommutative vortices has shown that the noncommutative version of $\varphi^{4} $-type models in two spatial dimensions with polynomial nonlinearities sustain non-collapsing soliton plateau and uncharged static vortex solutions, unlike what occurs in the commutative case. It has also been shown that the confining mechanism is provided by the $\star$-product which acts like a projection. Using this fact it has been further shown that large vortices can be produced by taking static suitable combinations of projections. In this process vortex-like structures with no angular dependence appear, and which have quantized radii  $R\sim\sqrt{n}$ for $n$
 integer. It is also known \cite{madore1},\cite{madore2} that spatial noncommutativity in 2-spheres in $ \mathbb R^{3}$ induces a lattice structure in the radial variable, where the quantized radii scale as $R\sim n$. Moreover this lattice can confine radial solutions.\\
On the other hand, recent work in non-linear optics (see {\it e.g.} \cite{peli}, \cite{wang}, \cite{chong}, \cite{cis} and \cite{kivs} for a review and background) has shown that optical lattices can trap coherent structures in the form of plateaus and vortices, with and without charge. The dynamics of these structures has been described successfully using modern asymptotics in terms of the Peierls-Nabarro (P-N) potential \cite{cis},\cite{panos} produced by the interaction between the coherent structure and the physical lattice.

The purpose of this letter is to show by means of an asymptotic analysis how the $\star$-product combined with the vortex structure induces a lattice in the spatial dimensions via a P-N like potential in the radial variable which, in turn, confines the vortex itself avoiding the collapse. We show, making use of a coherent state vortex-type solution with varying parameters, how the average Lagrangian \cite{whith} for the noncommutative Nonlinear Schr\"{o}dinger equation is equivalent to the known average Lagrangian for a vortex on a discrete lattice. The difference between the two Lagrangians being the actual form of the P-N potential. For the lattice the P-N potential is periodic with the lattice period, while on the noncommutative case considered here the P-N potential depends on $R^2$, which is the square of the vortex radius. This dependence gives a scaling $R\sim\sqrt{n}$ which is expected from the exact result described for the $\varphi^4$-type models.
We also exhibit how for vortices in two and three spatial dimensions, whose charge is of the same order of their large radius, the $\star$-product induces a P-N potential which results in an equispaced radial lattice. This shows how the splitting of ${\Bbb R}^3 $ in terms of fuzzy spheres placed on an equispaced radial lattice, as proposed in \cite{madore1},\cite{madore2}, arises naturally from
$\star$-product confining a vortex with large charge, large radius and small width. This results show how the dynamics of the coherent structures sustained by the noncommutativity can be described asymptotically in terms of the classical dynamics of coherent structures in lattices. Furthermore, the present analysis shows that the qualitative behavior for large vortices does not depend on the details of the noncommutative model chosen.

\section{Formulation}

We consider the Nonlinear Schr\"{o}dinger equation where the nonlinearity is given by the Moyal $\star$-product between the complex functions $u({\bf x},t)$,  $\bf x $ is the
2-dimensional spatial position and $t$ is the time.

The equation is given by the usual local Lagrangian \cite{whith}, suitably modified by the nonlocal product in the form:

\be \mathcal{L}=\int\int d\ {\bf x} dt [i(u_t\bar{u}-\bar{u_t}u)+|\nabla u|^2-\bar{u}\star u\star\bar{u}\star u ], \label{lagr1}\ee
where the Fourier transform of the $\star$-product  of two complex fields $u$ and $v$ in the Rieffel formula for the Moyal $\star$-product \cite{rieffel}  is given by:
\be \widehat{{u\star v}}({\bf k})=\frac{1}{(2\pi)^2}\int \hat{u}({\bf k}-{\bf p})\hat{v}({\bf p})e^{i({\bf k}-{\bf p})\wedge {\bf p}}d{\bf p}. \label{fourier} \ee
Here
$\hat{u}({\bf k},t)=\int e^{-i{\bf k}\cdot {\bf x}} u({\bf x},t)d{\bf  x} \label{fourier2}$ and ${\bf a}\wedge {\bf b} = \frac{\Theta}{2}(a_1,a_2)\left(
                                                              \begin{array}{cc}
                                                                0 & 1 \\
                                                                -1 & 0 \\
                                                              \end{array}
                                                            \right)\left(
                                                                     \begin{array}{c}
                                                                       b_1 \\
                                                                       b_2 \\
                                                                     \end{array}
                                                                   \right)$,
with $\Theta$ fixing the square of the length given by the noncommutativity of the spatial variables.

The equation associated with the Lagrangian (\ref{lagr1}) is the nonlocal Nonlinear Schr\"{o}dinger equation:
\be iu_t=\Delta u+2u\star \bar{u}\star u. \label{nls} \ee
This is the usual Schr\"{o}dinger equation  where the real potential $U(\bf x)$ is now dependent on the solution $\bar u\star u$.

In previous studies \cite{gopa1} of the Sine-Gordon equation, static (with no angular momentum) exact vortex-type solutions were found in the limit of $\Theta\rightarrow\infty$.

Analogous solutions of (\ref{nls}) can be obtained in the form
\be
u({\bf x}, t)=e^{i\sigma t} \zeta({\bf x}), \label{simpsol}
\ee
where $v$ is a real valued function of the position. Substitution into (\ref{nls}) readily gives
\be
-\sigma \zeta= \Delta \zeta +  \zeta\star \zeta \star \zeta. \label{3bis}
\ee
In the large noncommutativity limit (\ref{3bis}) becomes the same as the equation considered in \cite{gopa1}. Specially interesting solutions to  (\ref{3bis}) in this range of values of the
noncommutativity parameter discussed in the above cited work are given (using scaled variables) by:
\be
\varphi_n (r) = 2(-1)^n e^{-r^2} L_n (2r^2), \label{spsol}
\ee
with $r^2 = x^2 + y^2 $ and $\sigma=4\Theta$ and where $L_n$ are the Laguerre polynomials. It is known that sums of $\varphi_n (r)$ are also solutions. For example, taking
$$P_n(r) = \sum_{j=0}^n \varphi_j (r) \;\; \text{ and}\;\;\; W_n = \varphi_n (r) + \varphi_{n-1} (r). $$
Thus, using the well established fact that the Laguerre polynomials have caustics when $r^2 \sim n$ \cite{abra}, we obtain that the plateau $P_n(r)$ is practically flat up to $r \sim \sqrt n$ and confined to that region. On the other hand, $W_n$ has a peak at $r \sim \sqrt n$. Such a behavior
suggests trapping by an annular lattice with radii $R_n \sim \sqrt n$, for large $n$.

Moreover, it is known \cite{peli},\cite{kivs} both from analytical as well as numerical calculations that the usual Nonlinear Schr\"{o}dinger equation on the usual square lattice supports stable vortices due to the trapping of the vortex by the P-N potential, which prevents their collapse as it occurs in the limit of the continuum. We will show below, using the asymptotic analysis developed in
\cite{panos}, that the $\star$-product in equation (\ref{nls}) indeed generates, via the field, a P-N like potential responsible for the trapping of the vortex. We will further show how this P-N potential generates lattices with radii growing as $\sqrt n$ or as $n$, depending on the coherent state used for the averaging of the Lagrangian.\\


\section{Asymptotic solutions}
On the basis of the above considerations, we shall derive next an asymptotic solution to (\ref{nls}) in the form of a vortex with angular momentum (and charge one) given by the local behavior $r e^{i\theta}$
at the origin, and an envelope suggested by the exact solution with no angular momentum. Thus, we take as the coherent state trial function to average the Lagrangian (\ref{lagr1})
the expression
\be u(r,\theta,t)=a(t)re^{-(\frac{r-R(t)}{\omega(t)})^2}e^{i(\theta+\sigma)}e^{i(r-R(t))V(t)}, \label{ansatz} \ee
where the amplitude $a$, the width $\omega$ and the phase $\sigma$ are functions of time. The radius $R$ represents the location of the peak of the vortex and the velocity $V$ is the radial velocity of the maximum of the vortex.We will also consider charged
vortices with charge $m$ where the angular dependence is $r^m e^{im\theta}$ close to $r=0$.

Let us begin by studying the charge one vortex by substituting the trial function (\ref{ansatz}) into the Lagrangian (\ref{lagr1}) and performing the spatial integration we obtain an averaged Lagrangian for the parameters $a, \omega, R, V$
and $\sigma$.

 This Lagrangian is then varied and the modulation equations obtained for the parameters which give the approximate evolution of the vortex. The averaged Lagrangian has the form

\be \bar{\mathcal{L}}=\mathcal{L}_0+\mathcal{L}_*, \label{avgl}\ee
where $\mathcal{L}_0$ arises from the local terms in (\ref{lagr1}) and $\mathcal{L}_*$ is the contribution of the $\star$-product. We have, using the results in \cite{panos}, that

\be \frac{\mathcal{L}_0}{2\pi}=-2a^2\omega R^3\dot{\sigma}-2a^2R^3\omega(V\dot{R}-\frac{V^2}{2})-\frac{2a^2R^3}{3\omega}. \ee

To calculate $\mathcal{L}_*$ we begin by transforming the potential term with the $\star$-product in (\ref{lagr1}) into the Fourier space. We have, using the Parseval relation, that

\be \int\int\bar{u}\star u\star\bar{u}\star u \;d{\bf x}=\int\int|\bar{u}\star u|^2\;d{\bf x}=\int\int|\widehat{\bar{u}\star u}|^2({\bf k})\;d{\bf k}. \label{potential} \ee

The calculation of the last integrand is performed by making use of (\ref{fourier}) and the Fourier transform of coherent state (\ref{ansatz}) which is given by:
\be \hat{u}({\bf p})=a(t)e^{i\sigma(t)}\int_0^\infty\int_0^{2\pi}e^{ipr\cos(\theta-\varphi)}e^{i\varphi}re^{-(\frac{r-R}{\omega})^2}e^{i(r-R)V}rd\varphi dr, \label{fourier2} \ee
where ${\bf p}=p(\cos(\theta),\sin(\theta))$ and $\varphi$ is the polar angle of $\bf x$.

In the highly noncommutative limit the width $\omega$ is small. Hence the integrand in $r$ is peaked at $r=R$. Making the change of variables $r=R+\omega\xi$ we obtain, to leading order
in $\omega$,

\be \hat{u}({\bf p})=\omega a(t)e^{i\sigma(t)}e^{i\theta}R\int_{-R}^\infty\int_0^{2\pi}e^{ipR\cos\varphi}e^{i\varphi}e^{ik\omega\xi\cos\varphi} e^{-\xi^2}e^{i\omega\xi V}rd \varphi dr. \label{fourier3} \ee

Since we are interested in large vortices, we use the stationary phase approximation \cite{handels} on the angular integral. There are two stationary phase points at $\varphi=0$, $\varphi=\pi$ which give an oscillatory contribution. The radial integral is then calculated extending the lower limit to minus infinity to obtain:
\be \hat{u}({\bf p})=\frac{a\omega R^{3/2}}{(2\pi)^{3/2}p^{1/2}}\sin\left(pR+\frac{\pi}{4}\right)e^{-p^2\omega^2(1+V)^2}e^{i(\theta+\sigma)}. \label{ftfield1}\ee
In the same way we obtain
\be \hat{\bar u}({\bf p})=\frac{a\omega R^{3/2}}{(2\pi)^{3/2}\;p^{1/2}}\sin\left(pR+\frac{\pi}{4}\right)e^{-p^2\omega^2(1-V)^2}e^{-i(\theta+\sigma)}. \label{ftfield2}\ee
Consequently, using (\ref{ftfield1}) and (\ref{ftfield2}) together with (\ref{fourier}) allows us to arrive at an approximation for the integrand in (\ref{potential}) in the form

\be
\begin{split}
(\widehat{\bar{u}\star u})({\bf k})=a^2\omega^2R^3\int_0^\infty\int_0^{2\pi}e^{i\mathbf{k}\wedge\mathbf{p}}\frac{e^{-p^2\omega^2(1+V)^2}}{p^{1/2}}\frac{e^{-|\mathbf{k}-\mathbf{p}|^2\omega^2(1-V)^2}}{|\mathbf{k}-\mathbf{p}|^{1/2}}\\
\times\sin\left(pR+\frac{\pi}{4}\right)\sin\left(|\mathbf{k}-\mathbf{p}|R+\frac{\pi}{4}\right)e^{i\theta_1}e^{i\theta_2}pdpd\theta_1 d\theta_2\label{fourier4},\end{split}\ee
where $\theta_1$ and $\theta_2$ correspond to the angular coordinates of $\mathbf{p}$ and $\mathbf{k-p}$, respectively.
Again (\ref{fourier4}) will be evaluated approximately in the strongly noncommutative limit using the stationary phase method in the angular integral. To this end recall that both \textbf{k} and \textbf{p} are large since the vortex is narrow because $\omega$ is small. Using $p=\frac{q}{\omega}$ we obtain
\be \mathbf{k}\wedge\mathbf{p}=\frac{\Theta}{\omega}kq\sin\theta,\;\;
|\mathbf{p}-\mathbf{k}|=\sqrt{\frac{q^2}{w^2}-2\frac{q}{\omega}k\cos\theta+k^2}\ee

Since in the strongly noncommutative limit $\textbf{k}$ is also large we know that the points of the stationary phase are for $\theta =\pi/2$ and $\theta=3\pi/2$. Using again the same type of calculation as in the derivation of equation (\ref{fourier3}) we obtain:

\be\begin{split} (\widehat{\bar{u}\star u})({\bf k})=\omega^{1/2}e^{-\omega^2k^2}\;\frac{a^2\omega^2}{(2\pi)^3}
\left(\frac{\omega}{\Theta}\right)^{1/2}R\int_0^\infty\sin \left (kq\frac{\Theta}{\omega}\right )e^{-q^2(1+V)^2}\\
\times\sin\left (\frac{R}{\omega}q\right)\sin \left( R\sqrt{\frac{q^2}{\omega^2}+k^2}\right )\frac{e^{-(q^2+k^2\omega^2)(1-V)^2}}{\sqrt{\frac{q^2}{\omega^2}+k^2}}q^{1/2}dq.\end{split}\ee

This integral is again evaluated asymptotically for larger $R$ using the method of stationary phase. We need to observe that for small $\omega$ and to leading order of this width the stationary phase point is $q=\frac{k\omega}{2}$. We then have

\begin{align}(\widehat{\bar{u}\star u})({\bf k})=&\frac{R^{5/2}}{(2\pi)^3\omega^{1/2}}e^{-\omega^2 k^2}\;a^2\omega^2 \left(\frac{\omega}{\Theta}\right)^{1/2}\nonumber\\&\times\sin\left (\frac{k^2}{2}\Theta\right)\sin^2(kR)e^{-\frac{\omega^2}{4}k^2
((1+V)^2+\frac{5}{4}(1-V)^2)}.\label{18}\end{align}

Finally, integrating this last expression over $\bf k$ we obtain that the Lagrangian term $\mathcal{L}_*$ is given by:
\be \mathcal{L}_*=\frac{a^4R^{5/2}\omega^2}{(2\pi)^3\Theta}\int_0^\infty e^{-\omega^2 k^2[2+\frac{1}{2}(1+V)^2+\frac{5}{8}(1-V)^2]}\;\sin^2 \left(\frac{k^2}{2}\Theta\right)\sin^4(kR)kdk,\ee
and, evaluating once more with the method of stationary phase for large $R$ we obtain:

\be \mathcal{L}_*=-\frac{a^4}{(8\pi)^{3}\Theta}\;\omega^2 R^{5/2}\left( \frac{1}{2\omega^2} +F(R,\omega,V)\right),\ee
where the function $F$ is the analogue of the Peierls-Nabarro potential for the lattice generated by the noncommutative self interaction of the field, and takes the form:

\be F(R,\omega,V)=\frac{1}{4R^{1/2}}\cos\left(\frac{R^2}{2\Theta}+\frac{\pi}{4}\right)e^{-\omega^2R^2[2+\frac{(1+V)^2}{2}+\frac{5(1-V)^2}{8}]}.\label{21}\ee

Hence the final average Lagrangian is:

\begin{align} \mathcal{L}=&2a^2\omega R^3\dot{\sigma}+2a^2\omega R+2a^2\omega R^3(V\dot{R}-\frac{V^2}{2})\nonumber
\\&+2\frac{a^2R^3}{3\omega}-\frac{a^4 R}{(8\pi)^3\Theta}-F(R,\omega,V). \label{finallagr}\end{align}

This Lagrangian is, except for the $R^2$ dependance in the potential, the same average Lagrangian obtained in [cite] for large Nonlinear Schr\"{o}dinger commutative vortices on a discrete lattice. However in the present case the lattice is generated by the $\star$-product and it manifests itself in the $R^2$ dependance of the Peierls-Nabarro potential.

\section{Vortex solutions and radial stability}

The approximate dynamics of the vortex is obtained from the variational equations of (\ref{finallagr}). These are

\begin{eqnarray}
  \delta\sigma&:& \frac{d}{dt}(2a^2\omega R)=0, \nonumber\\
  \delta a&:& 4a\omega R^3\frac{d\sigma}{dt}+\frac{4aR^3}{3\omega}-\frac{Ra^3}{2\Theta}+\partial_a F=0, \nonumber\\
  \delta\omega&:& 2a^2R^3 \frac{d\sigma}{dt}-\frac{2a^2R^3}{3\omega^2}+\partial_\omega F=0, \label{vareq}\\
  \delta V&:& \dot{R}-V-\partial_V F=0, \nonumber\\
  \delta R&:& \frac{d}{dt}2a^2\omega R^3V+\partial_R \mathcal{L}=0.\nonumber
\end{eqnarray}
The last expression above is the equation of motion for the peak of the vortex, analogous to a particle in the Peierels-Nabarro potential.

The dynamics of the solutions to (\ref{vareq}) simplifies for large vortices in lattices, with low kinetic energy, {\it i.e.} for $V\ll1$. In this case, since $\omega$ is small, the dominating terms for the $\delta a$ and $\delta\omega$ equations can be readily solved to give a steady vortex amplitude width relation in the form

\be a=32\sqrt{\frac{\Theta}{3\omega}}.\ee

For large vortices with $R\omega$ still small and with small kinetic energy $V\ll1$ the dynamics of the peak is described by the simple equation

\begin{eqnarray}
\frac{dR}{dt}&=&V,\nonumber\\
a^2\omega\frac{d(RV)}{dt}&=&\frac{a^4\omega^4}{2\Theta^2}R^{3}\sin\left(\frac{R^2}{2\Theta}+\frac{\pi}{4}\right).\label{simplevortex}\end{eqnarray}

Equation (\ref{simplevortex}) shows that the vortex moves in the lattice which was generated by the $\star$-product. The fixed points are the possible equilibrium positions and are given by
\be R_n=\sqrt{2n\pi\Theta}.\ee
The odd values of $n$ give stable vortices, while the even values of $n$ give instability. If a vortex starts at an unstable value it will shrink radiating until it is trapped at the lower minimum of the potential. It is to be noted that the scaling of the radius is the same as the one obtained by the exact trapped solutions of $\varphi^4$-type noncommutative models.
This result has a simple interpretation in terms of the Landau cells used in \cite{madore1}. In fact the P-N  potential generated by the $\star$-product induces an annular lattice, where the $n$-th annulus has an area $ A_n =2\pi R_n(R_{n+1} -R_n)$
and where $R_n$ is given by (\ref{simplevortex}). As $n\rightarrow\infty$ so that $A_n \sim \pi^2 \Theta$, which gives the constraint area of the Landau cell. We thus can say that the lattice induced by the $\star$-product is a lattice of Landau cells.
The same calculation when performed for a charge $m$ vortex, with $R$ of the same order as $m$, gives an equispaced lattice. In fact the asymptotic evolution of the integrals in Eq.(\ref {fourier3}) replaces the term $\sin^2 (kR)$ of (\ref{18}) by
$\sin^2 (k + \frac{1}{k})R$. This results in a P-N potential of the form (\ref{21}) where the term $(\cos\frac{R^2}{2\theta} +\frac{\pi}{4})$ is replaced by $\cos (R +\frac{\pi}{4})$. this induces an equispaced lattice. The area of the Landau cells is again constant, as can be readily verified. Finally, when we go to ${\mathbb R}^3$ and use in the averaging of the Lagrangian coherent structures with no angular dependence, we obtain
shell lattices with $R_n \sim \sqrt n$; while when including an angular dependence given by the spherical function
$Y_{lm}(\theta,\varphi)$,
induces - for large values of $l$ and as a consequence of the form of the coherent state - equispaced lattices, thus recovering in a natural way the noncommutative structure proposed in \cite{madore1}.

\section{Discussion and Conclusions}

We have shown that the effect of noncommutativity of the spatial variables, when averaged on the appropriate coherent vortex or plateau-like states, induces an effective spatial lattice of Landau cells whose distribution and sizes depend on the coherent states in question. This shows that the effect of noncommutativity on coherent structures whose local width is comparable to the spatial scale $\Theta$ of the $\star$-product behave as classical structures on a physical lattice and allows us to calculate the lattice and the corresponding dynamics of the noncommutative coherent structure. It is to be remarked that the lattice structures in three space dimensions have been constructed using a group on the sphere and taking the Casimir values as the possible values of the radii inducing a lattice. We have shown that in the appropriate coherent states the $\star$-product induces the same lattice in the radial variable.

We also observe that unlike physical lattices which are not translation invariant, the lattices induced by the $\star$-product are translation invariant. Because of this reason coherent states can move with uniform velocity. We thus see that the effect of the $\star$-product has no classical analogy since it is capable of forming a lattice to support the coherent structure without loosing the uniform translational motion.

We end by remarking that in the weak noncommutative limit, which is the opposite limit to the one considered here, equation (\ref{lagr1}) resembles a high order non-linear system of equations which incorporates Raman scattering effects \cite{kivs}. However at the present time the existence of an optical analogue of equation (\ref{lagr1}) in the strongly noncommutative limit is not known.

\end{document}